\documentclass[11pt]{article} 
\usepackage{epsfig} 
\usepackage{cite}  
\newcommand{\triangleNLO}[3]{
\mbox{\parbox{3cm}{\hspace{0.25cm}
\begin{picture}(2.5,1.4)
\thicklines
\put(0.3,0.7){\vector(1,0){0.1}}
\put(1.7,0.2){\vector(1,0){0.1}}
\put(1.7,1.2){\vector(1,0){0.1}}
\put(0,0.7){\line(1,0){0.5}}
\put(1,1.2){\line(0,-1){1}}
\put(1,1.2){\line(1,0){1}}
\put(1,0.2){\line(1,0){1}}
\put(1,0.7){\circle{1}}
\put(0.25,0.9){\makebox(0,0)[b]{$#1$}}
\put(2.05,1.2){\makebox(0,0)[l]{$#2$}}
\put(2.05,0.2){\makebox(0,0)[l]{$#3$}}
\end{picture}
}}
\hfill}

\newcommand{\bubbleNLO}[1]{
\mbox{\parbox{2.5cm}{\hspace{0.25cm}
\begin{picture}(2,1)
\thicklines
\put(0.3,0.5){\vector(1,0){0.1}}
\put(0,0.5){\line(1,0){2}}
\put(1,0.5){\circle{1}}
\put(0.25,0.7){\makebox(0,0)[b]{$#1$}}
\end{picture}
}}
\hfill}

\newcommand{\trianglecrossNLO}[3]{
\mbox{\parbox{3cm}{\hspace{0.25cm}
\begin{picture}(2.5,1.4)
\thicklines
\put(0.3,0.7){\vector(1,0){0.1}}
\put(1.9,0.2){\vector(1,0){0.1}}
\put(1.9,1.2){\vector(1,0){0.1}}
\put(0,0.7){\line(1,0){0.5}}
\put(0.5,0.7){\line(1,1){0.5}}
\put(0.5,0.7){\line(1,-1){0.5}}
\put(1.5,1.2){\line(-1,-2){0.5}}
\put(1.0,1.2){\line(1,-2){0.18}}
\put(1.5,0.2){\line(-1,2){0.18}}
\put(1,1.2){\line(1,0){1}}
\put(1,0.2){\line(1,0){1}}
\put(0.25,0.9){\makebox(0,0)[b]{$#1$}}
\put(2.05,1.2){\makebox(0,0)[l]{$#2$}}
\put(2.05,0.2){\makebox(0,0)[l]{$#3$}}
\end{picture}
}}
\hfill}

\newcommand{\bubbleLO}[1]{
\mbox{\parbox{2.5cm}{\hspace{0.25cm}
\begin{picture}(2,1)
\thicklines
\put(0.3,0.5){\vector(1,0){0.1}}
\put(0,0.5){\line(1,0){0.5}}
\put(1.5,0.5){\line(1,0){0.5}}
\put(1,0.5){\circle{1}}
\put(0.25,0.7){\makebox(0,0)[b]{$#1$}}
\end{picture}
}}
\hfill}

\def\e{\epsilon}
\def\d{\hbox{d}}

\def\gam{\Gamma}

\begin{document} 
\unitlength1cm 
\begin{titlepage} 
\vspace*{-1cm} 
\begin{flushright} 
ZU--TH 12/05\\
hep-ph/0507061\\
July 2005
\end{flushright} 
\vskip 3.5cm 

\begin{center} 
{\Large\bf Two-Loop Quark and Gluon Form Factors\\[2mm]
 in Dimensional Regularisation}
\vskip 1.cm 
{\large  T.~Gehrmann}, {\large T.~Huber} 
and {\large D.~Ma\^{\i}tre}
\vskip .7cm 
{\it Institut f\"ur Theoretische Physik, Universit\"at Z\"urich,
Winterthurerstrasse 190,\\ CH-8057 Z\"urich, Switzerland} 
\end{center} 
\vskip 2cm 

\begin{abstract} 
We compute the two-loop corrections to the massless quark form factor 
$\gamma^* \to q\bar q$ and gluon form factor $H\to gg$ to all orders 
in the dimensional regularisation parameter $\e=(4-d)/2$. The 
two-loop contributions to the form factors are reduced to
linear combinations of master integrals, which are computed 
in a closed form, expressed as $\Gamma$-functions and generalised 
hypergeometric functions of unit argument. Using the newly
developed {\tt HypExp}-package, these can be expanded to any desired order, 
yielding Laurent expansions in $\e$. We provide expansions 
of the form factors to order $\e^2$, as required for 
ultraviolet renormalisation and
infrared factorisation of the three-loop form factors.

\end{abstract} 
\vfill 
\end{titlepage} 
\newpage 

\section{Introduction}
The infrared pole structure of renormalised multi-loop amplitudes 
in dimensional regularisation with $d=4-2\e$ space-time dimensions can be 
predicted from an infrared factorisation formula, which was first 
conjectured in~\cite{catani}, where it was formulated up to two loops. A 
proof of the formula, together with an explicit formulation up to three
loops was derived later in~\cite{sterman}. The simplest multi-loop amplitudes
where the infrared factorisation formula can be applied are three-point 
functions, involving two partons coupled to an external current: the 
quark form factor $\gamma^*\to q\bar q$ and the gluon form factor 
$H\to gg$. The QCD corrections to these form factors can in particular be 
used to fix a priori unknown constants in the infrared factorisation formula,
thus enabling an unambiguous prediction for multi-loop amplitudes involving more
than two external partons. 

In the infrared factorisation formula for a given form factor (or 
more generally for a given multi-leg amplitude)
at a certain number of loops, infrared singularity operators act on the 
form factor evaluated with a lower number of loops. The 
infrared singularity operators contain explicit infrared poles
$1/\e^2$ and $1/\e$. They 
do therefore project subleading terms in $\e$ from the lower order 
form factors. 

At present, two-loop corrections to the massless quark~\cite{vanneerven} and 
gluon~\cite{harlander} form
factors are known 
to order $\e^0$. Two-loop corrections to this 
order were also obtained for massive quarks~\cite{breuther}. The 
infrared structure of the massless form factors and infrared 
cancellations with real radiation contributions are described in detail 
in~\cite{irstruc}. Very recently, results to 
order $\e^2$ were obtained for the quark form factor~\cite{mvvnew}.

The calculation of
these corrections proceeds through a 
reduction~\cite{chet,laporta,gr,babis} of all two-loop Feynman 
integrals appearing in the form factors to a small set of master integrals. 
The reduction is exact in $\e$, such that the evaluation of the form factors 
is limited only by the order to which the master integrals can be computed.
The massless two-loop form factors contain three two-loop master integrals,
which can be computed either using various analytical methods~\cite{smirnov}
or numerically order-by-order in their Laurent expansion using the 
sector decomposition algorithm~\cite{secdec}.
Up to now, exact expressions were known only for two of these master 
integrals, while the third (the so-called two-loop crossed triangle 
graph) was known only 
as a Laurent expansion up to finite terms~\cite{kl}.

In this letter, we derive an exact expression for the two-loop crossed
 triangle graph in terms of generalised 
hypergeometric functions of unit argument
in Section~\ref{sec:twol}. Using 
the {\tt HypExp}-package~\cite{hypexp} for the Laurent expansion of 
 generalised hypergeometric functions, this can be expanded to any 
desired order in $\e$.
 Together with the exact expressions for the 
one- and two-loop quark and gluon form factors in Section~\ref{sec:ff}, 
this allows the expansion of  these form factors to higher orders in 
$\e$. For illustration, we list the one-loop form factors to 
order $\e^4$ and the two-loop form factors to order $\e^2$ in 
Section~\ref{sec:ffexp}; these orders appear for example in the 
infrared factorisation of the corresponding three-loop from factors. 
Finally, Section~\ref{sec:conc} contains conclusions and an outlook.

\section{Two-loop master integrals}
\label{sec:twol}
The virtual two-loop vertex master integrals were first derived
to order $\e^0$ 
in~\cite{kl} in the context of the calculation of the two-loop quark form 
factor~\cite{vanneerven}. All but the crossed triangle graph $A_6$
can be expressed in terms of $\Gamma$-functions to all orders in $\e$.

Factoring out a common
\begin{equation}
S_\Gamma = \left(\frac{(4\pi)^{\e}}{16\pi^2\,\Gamma(1-\e)
}\right)\, ,
\end{equation}
and introducing $q^2 = (p_1+p_2)^2$, they read
\begin{eqnarray}
A_{2,{\rm LO}} &=& \bubbleLO{p_{12}} \nonumber \\
&=&
\int \frac{\d^d k}{(2\pi)^d} \,  
\frac{1}{k^2(k-p_1-p_2)^2 }
\nonumber \\ &=&S_\Gamma \,\left(-q^2\right)^{-\e}\,
 \frac{ \Gamma(1+\e)\Gamma^3(1-\e) }{
 \Gamma(2-2\e)}\frac{i}{\e} \;,
\\
A_3 &=& \bubbleNLO{p_{12}} \nonumber \\
&=&\int \frac{\d^d k}{(2\pi)^d} \,\int \frac{\d^d l}{(2\pi)^d} \, 
\frac{1}{k^2 l^2 (k-l-p_1-p_2)^2}
\nonumber \\ 
&=& S_\Gamma^2\,  \left( -q^2 \right)^{1-2\e}\, 
\frac{\Gamma(1+2\e)\Gamma^5(1-\e)}{
\Gamma(3-3\e)} 
\frac{-1}{2(1-2\e)\e}\;,
\\
A_4 &=& \triangleNLO{p_{12}}{p_1}{p_2} \nonumber \\ &=&
\int \frac{\d^d k}{(2\pi)^d} \,\int \frac{\d^d l}{(2\pi)^d} \, 
\frac{1}{k^2 l^2 (k-p_1-p_2)^2 (k-l-p_1)^2}
\nonumber \\ 
&=& S_\Gamma^2 \,\left(-q^2\right)^{-2\e}\,
 \frac{ \Gamma(1-2\e)\Gamma(1+\e)\Gamma^4(1-\e) 
 \Gamma(1+2\e)}{
\Gamma(2-3\e)}\frac{-1}{2(1-2\e)\e^2} \;.
\end{eqnarray}
No exact expression for $A_6$ was known up to now. 
Following the steps outlined in~\cite{kl}, we obtain
\begin{eqnarray}
A_6 &=& \trianglecrossNLO{p_{12}}{p_1}{p_2} \nonumber \\ &=&
\int \frac{\d^d k}{(2\pi)^d} \,\int \frac{\d^d l}{(2\pi)^d} \, 
\frac{1}{k^2 l^2 (k-p_1-p_2)^2 (k-l)^2 (k-l-p_2)^2 (l-p_1)^2}
\nonumber \\ 
&=&  S_\Gamma^2 \,\left(-q^2\right)^{-2-2\e}\, \left[ -\,\frac{\gam^3(1-\e)\,\gam(1+\e)\,\gam^4(1-2\e)\,\gam^3(1+2\e)}{\e^4\,
\gam^2(1-4\e)\,\gam(1+4\e)}\right.\nonumber \\
&& +\,\frac{\gam^4(1-\e)\,\gam(1+\e)\,\gam(1-2\e)\,\gam(1+2\e)}{2\,\e^4\,\gam(1-3\e)}       \, \,
_3F_2(1,-4\e,-2\e;1-3\e,1-2\e;1)\nonumber \\
&& -\,\frac{4\,\gam^4(1-\e)\,\gam(1-2\e)\,\gam(1+2\e)}{\e^2\,(1+\e)\,(1+2\e)\,\gam(1-4\e)} \, \,
_3F_2(1,1,1+2\e;2+\e,2+2\e;1)\nonumber \\
&&\left.-\,\frac{\gam^5(1-\e)\,\gam(1+2\e)}{2\,\e^4\,\gam(1-3\e)} \, \, _4F_3(1,1-\e,-4\e,-2\e;1-3\e,1-2\e,1-2\e;1)\right]\;.
\end{eqnarray}

While $A_{2,LO}$, $A_3$ and $A_4$ can be expanded using any standard 
computer algebra programme, the expansion of $A_6$ requires 
the expansion of generalised hypergeometric functions in their parameters. 
For this purpose, a dedicated package, {\tt HypExp}~\cite{hypexp}, was 
developed recently. Using this, we obtain the  eighth-order expansion:
\begin{eqnarray}
A_6 &=&  S_\Gamma^2 \,\left(-q^2\right)^{-2-2\e}\, \Bigg[
 -\frac{1}{\e^4} + 
\frac{5\pi^2}{6\e^2} + \frac{27}{\e}\zeta_3 + \frac{23\pi^4}{36} 
+ \left(117\zeta_5-8\pi^2\zeta_3\right)\e\nonumber \\
&&+ \left(\frac{19\pi^6}{315}-267\zeta^2_3\right)\e^2 + \left(-\frac{109\pi^4}{10}\zeta_3-40\pi^2\zeta_5 - 6\zeta_7\right)\e^3\nonumber \\
&&+ \left(44\pi^2\zeta_3^2-\frac{1073\pi^8}{3024}-2466\,\zeta_3\,\zeta_5+264\zeta_{5,3}\right)\e^4 + {\cal O}(\e^5) \Bigg]\; ,
\end{eqnarray}
where we encountered a multiple zeta value in the last term.

\section{Quark and gluon form factors at two loops}
\label{sec:ff}

The tree-level quark and gluon form factors are obtained by normalising the 
corresponding tree-level vertex functions to unity:
\begin{equation}
F^{(0l)}_q = 1 \, , \qquad F^{(0l)}_g = 1\,.
\label{eq:ffnorm}
\end{equation}
The unrenormalised one-loop and two-loop form factors are calculated from 
the relevant Feynman diagrams. Using integration-by-parts~\cite{chet} and 
Lorentz invariance~\cite{gr} identities (which can be solved 
symbolically for massless two-loop vertex integrals, see the appendix 
of~\cite{crossed}), these can be 
reduced~\cite{laporta,gr,babis} to the master integrals listed in 
Section~\ref{sec:twol}.

The unrenormalised one-loop quark and gluon form factors read:
\begin{eqnarray}
F^{(1l,B)}_q &=&  -i g^2 \, \frac{N^2-1}{N} \, \frac{d^2-7d+16}{2\,(d-4)}
\, A_{2,{\rm LO}}\,,\\
F^{(1l,B)}_g &=&  i g^2 \, N \,\frac{d^3-16d^2+68d-88}{(d-4)\,(d-2)}
\, A_{2,{\rm LO}}\,,
\end{eqnarray}
where $N=3$ is the number of colours  and $g$ is the bare QCD coupling 
parameter.

The unrenormalised two-loop quark and gluon form factors for 
$N_F$ massless quark flavours are:
\begin{eqnarray}
\lefteqn{F^{(2l,B)}_q = g^4 \, \frac{N^2-1}{N}\bigg\{ 
-\frac{N^2-1}{N} \, \frac{(d^2-7d+16)^2 }{4\,(d-4)^2} \, 
A^2_{2,{\rm LO}} }\nonumber \\ 
&& +N \,\frac{(d^5-18d^4+138d^3-552d^2+1144d-980)\,(3d-8)}{2\,
(d-3)\,(d-4)^3} \, \frac{A_3}{q^2} \nonumber \\
&& +\frac{1}{N}\,
\frac{(9 d^6-358 d^5+4309 d^4-24466 d^3+72896 d^2-110064 d+66080)
\,(3 d-8)}{16(d-3) (d-4)^3 (2 d-7)}\, \frac{A_3}{q^2} \nonumber \\
&& + N \, \frac{
3d^6-82d^5+819d^4-4030d^3+10344d^2-12824d+5632}
{4\,(d-1)\,(d-4)^2\,(3d-8)}
\,A_4 \nonumber \\
&& -\frac{1}{N}\, \frac{(21d^6-789d^5+9422d^4-53864d^3+163200d^2-253472
d+159232)}{16\,(3d-8)\,(2d-7)\,(d-4)^2}\, A_4\nonumber \\
&& +N_F\,\frac{(3d^3-31d^2+110d-128)\,(d-2)}{2\,(d-1)\,(d-4)\,(3d-8)}\, 
A_4 \nonumber \\
&& -\frac{1}{N}\, \frac{d^3-20d^2+104d-176}{32\,(2d-7)}
(q^2)^2\,A_6 \bigg\}\;,\label{eq:fqunren}\\
\lefteqn{F^{(2l,B)}_g = g^4 \, \bigg\{-
N^2 \, \frac{(d^3-16d^2+68d-88)^2 }{(d-4)^2\,(d-2)^2} \, 
A^2_{2,{\rm LO}}  }\nonumber \\ &&
-N^2 \,\frac{1}{2\,
(d-1)\,(d-2)^2\,(d-3)\,(d-4)^3\,(2d-5)\,(2d-7)} \, 
\Big(192 d^{10}-6947 d^9
\nonumber \\ && \hspace{1cm}
+105470d^8
-907248d^7+4958664d^6
-18113645d^5+44930982d^4\nonumber \\ && \hspace{1cm}-74791460d^3
+79854504d^2
-49204128d+13194496
\Big)\,\frac{A_3}{q^2} \nonumber \\
&& - N\, N_F \frac{2d^6-45d^5+377d^4-1610d^3+3868d^2-5136d+3008}{
(d-1)\,(d-2)\,(d-3)\,(d-4)^2}\,\frac{A_3}{q^2}\nonumber \\
&& +\frac{N_F}{N}\,
\frac{1}{4\,(d-2)\,
(d-3)\,(d-4)^2\,(2d-5)\,(2d-7)}\, 
\Big(70d^7-1663d^6+16290d^5\nonumber \\ && \hspace{1cm}
-86031d^4+266004d^3-483356d^2+479360d-200704
\Big)\,\frac{A_3}{q^2} \nonumber \\
&& - N^2 \, \frac{1}
{2\,(d-1)\,(d-2)^2\,(d-4)^2\,(2d-5)\,(2d-7)}\,\Big( 
108 d^8-2661 d^7+28822d^6\nonumber \\ && \hspace{1cm}-177546d^5+
674735d^4-1607602d^3+2325996d^2-1848920d+607968
\Big)
\,A_4 \nonumber \\
&& +N \,N_F\, \frac{2d^4-28d^3+130d^2-228d+104}
{(d-1)\,(d-2)\,(d-4)}\, A_4\nonumber \\
&& -\frac{N_F}{N}\, \frac{(46d^4-545d^3+2395d^2-4606d+3248)\,(d-6)}
{4\,(d-2)\,(d-4)\,(2d-5)\,(2d-7)}\, A_4\nonumber \\
&& -N^2\, \frac{3\,(3d-8)\,(d-3)}{4\,(2d-5)\,(2d-7)}
(q^2)^2\,A_6 \nonumber \\ 
&& -\frac{N_F}{N} \, \frac{(d-4)\,(2d^3-25d^2+94d-112)}
{8\,(d-2)\,(2d-5)\,(2d-7)}
(q^2)^2\,A_6
\bigg\}\;.
\end{eqnarray}
The renormalised form factors are obtained by introducing  the
renormalised  QCD coupling constant and the renormalised effective 
coupling of $H$ to the gluon field strength~\cite{harlander}, and subsequent 
expansion in powers of the renormalised coupling.

\section{Expansion of two-loop form factors}
\label{sec:ffexp}

The renormalised form factors are expanded in the renormalised coupling 
constant. 
In the $\overline{{\rm MS}}$ scheme,
the bare coupling $\alpha_0 = g^2/(4\pi)$ is related to the 
 renormalised coupling
$\alpha_s\equiv \alpha_s(\mu^2)$,
evaluated at the renormalisation scale $\mu^2$ by 
\begin{equation}
\alpha_0\mu_0^{2\e} S_\e = \alpha_s \mu^{2\e}\left[
1- \frac{11 N - 2 N_F}{6\e}\left(\frac{\alpha_s}{2\pi}\right)
+{\cal O}(\alpha_s^2) \right]\; ,
\end{equation}
where
\begin{displaymath}
S_\e =(4\pi)^\e e^{-\e\gamma}\qquad \mbox{with the Euler constant }
\gamma = 0.5772\ldots
\end{displaymath}
and $\mu_0^2$ is the mass parameter introduced
in dimensional regularisation to maintain a
dimensionless coupling
in the bare QCD Lagrangian density. For simplicity, we set $\mu^2 = q^2$.
If the squared momentum transfer $q^2$ is space-like ($q^2<0$), the form 
factors are real, while they acquire imaginary parts for time-like 
$q^2$. These imaginary parts (and corresponding real parts) arise from 
the $\e$-expansion of 
\begin{equation}
\Delta(q^2) = (-\mbox{sgn}(q^2) - i0)^{-\e}\;.
\end{equation}

The renormalised form factors can then be written as
\begin{equation}
{F}_{q,g}(q^2) = 1 + \left(\frac{\alpha_s}{2\pi}\,\Delta(q^2) \right) 
F^{(1)}_{q,g} + 
 \left(\frac{\alpha_s}{2\pi}\,\Delta(q^2) \right)^2 F^{(2)}_{q,g} 
+ {\cal O} (\alpha_s^3)\;.
\end{equation}

Expanding the first and second order coefficients of the form factors 
to $\e^4$ and $\e^2$ respectively, we obtain:
\begin{eqnarray}
F^{(1)}_{q} &= & \left( N-\frac{1}{N}\right)
  \Bigg[ -\frac{1}{2\e^2} - \frac{3}{4\e} - 2 + 
\frac{\pi^2}{24} 
+ \left( -4 + \frac{\pi^2}{16} + \frac{7}{6}\zeta_3 \right)
\e \nonumber \\
&& 
+ \left( - 8
          + \frac{\pi^2}{6}
          + \frac{7}{4}\zeta_3
          + \frac{47\pi^4}{2880} \right) \e^2\nonumber \\ &&  
+\left(           - 16
          - \frac{7\pi^2}{72}\zeta_3
          + \frac{14}{3}\zeta_3
          + \frac{31}{10}\zeta_5
          + \frac{\pi^2}{3}
          + \frac{47\pi^4}{1920}
\right) \e^3 \nonumber \\ &&
+\left(         - 32
          - \frac{7\pi^2}{48}\zeta_3
          + \frac{28}{3}\zeta_3
          - \frac{49}{36}\zeta_3^2
          + \frac{93}{20}\zeta_5
          + \frac{2\pi^2}{3}
          + \frac{47\pi^4}{720}
          + \frac{949\pi^6}{241920}
 \right) \e^4\Bigg]
+ {\cal O}(\e^5) \;, \\
F^{(1)}_{g} &= & N \Bigg[ -\frac{1}{\e^2} 
- \frac{11}{6\e}  + 
\frac{\pi^2}{12} 
+ \left( -1 + \frac{7}{3}\zeta_3 \right)
\e 
+ \left( - 3
+ \frac{47\pi^4}{1440} \right) \e^2  \nonumber \\ &&
+ \left(
          - 7
          - \frac{7\pi^2}{36}\zeta_3
          + \frac{31}{5}\zeta_5
          + \frac{\pi^2}{12}
\right) \e^3 \nonumber \\ &&
+ \left( 
          - 15
          + \frac{7}{3}\zeta_3
          - \frac{49}{18}\zeta_3^2
          + \frac{\pi^2}{4}
          + \frac{949\pi^6}{120960}
\right) \e^4
\Bigg] + 
\frac{N_F}{3\e} + {\cal O}(\e^5) , \\
F^{(2)}_{q} &= & \left( N-\frac{1}{N}\right)\;
  \Bigg\{ \; N \Bigg[ \frac{1}{8\e^4} + \frac{17}{16\e^3}
+ \frac{433}{288\e^2} 
\nonumber \\&&
+ \frac{1}{\e} \left( \frac{4045}{1728} -\frac{11\pi^2}{96} 
+\frac{7}{24}\zeta_3 \right)
+\left(-\frac{9083}{10368}
          - \frac{521\pi^2}{1728}
          + \frac{13}{18}\zeta_3           + \frac{23\pi^4}{2880}
   \right) \nonumber \\ &&
+ \left(          - \frac{1244339}{62208}
          - \frac{11\pi^2}{48} \zeta_3 
          + \frac{4235}{432} \zeta_3
          + \frac{163}{40} \zeta_5
          - \frac{10427\pi^2}{10368}
          + \frac{29\pi^4}{1440} 
\right) \e \nonumber \\ &&
+ \bigg(          - \frac{36528395}{373248}
          - \frac{77\pi^2}{432} \zeta_3
          + \frac{109019}{2592} \zeta_3
          - \frac{403}{72} \zeta_3^2
          + \frac{529}{30} \zeta_5\nonumber \\ && \hspace{1cm}
          - \frac{181451\pi^2}{62208}
          + \frac{8759\pi^4}{51840} 
          + \frac{47\pi^6}{7560} 
\bigg) \e^2 
\Bigg]\nonumber \\
&&+\frac{1}{N} \Bigg[ -\frac{1}{8\e^4} - \frac{3}{8\e^3}
+ \frac{1}{\e^2} \left( -\frac{41}{32} +\frac{\pi^2}{48} \right)
\nonumber \\&&
+ \frac{1}{\e} \left( -\frac{221}{64} 
+\frac{4}{3}\zeta_3 \right)
+\left(-\frac{1151}{128}
          - \frac{17\pi^2}{192}
          + \frac{29}{8}\zeta_3   + \frac{13\pi^4}{576}
   \right) \nonumber \\ &&
+\left(          - \frac{5741}{256}
          - \frac{7\pi^2}{18} \zeta_3
          + \frac{839}{48} \zeta_3
          + \frac{23}{10} \zeta_5
          - \frac{71\pi^2}{128} 
          + \frac{19\pi^4}{320} 
\right) \e\nonumber \\ &&
+\left(   - \frac{27911}{512}
          - \frac{9\pi^2}{16} \zeta_3
          + \frac{6989}{96} \zeta_3
          - \frac{163}{9} \zeta_3^2
          + \frac{231}{40} \zeta_5
          - \frac{613\pi^2}{256} 
          + \frac{3401\pi^4}{11520} 
          - \frac{223\pi^6}{17280} 
\right) \e^2
\Bigg]\nonumber \\
&&+ N_F \Bigg[ -\frac{1}{8\e^3} - \frac{1}{18\e^2}
+ \frac{1}{\e} \left( \frac{65}{432} +\frac{\pi^2}{48} \right)
+ \left( \frac{4085}{2592} +\frac{23\pi^2}{432} 
+\frac{1}{36}\zeta_3 \right)\nonumber \\ &&  
+\left(          \frac{108653}{15552}
          - \frac{119}{108} \zeta_3
          + \frac{497\pi^2}{2592}
          + \frac{\pi^4}{1440} 
\right) \e\nonumber \\ &&
+\left(           \frac{2379989}{93312}
          - \frac{5\pi^2}{54} \zeta_3
          - \frac{3581}{648} \zeta_3
          - \frac{59}{60} \zeta_5
          + \frac{9269\pi^2}{15552} 
          - \frac{145\pi^4}{10368} 
\right) \e^2
\Bigg]
 \Bigg\}+ {\cal O}(\e^3) 
\;, \label{eq:quarkff} \\
F^{(2)}_{g} &= &   N^2 \,\Bigg[ \frac{1}{2\e^4} + \frac{77}{24\e^3}
+ \frac{1}{\e^2} \left(\frac{175}{72}-\frac{\pi^2}{24}\right) 
\nonumber \\&&
+ \frac{1}{\e} \left(
          - \frac{119}{54}
          - \frac{25}{12} \zeta_3
          - \frac{11\pi^2}{144}
 \right)
+\left(
            \frac{8237}{648}
          - \frac{33}{4} \zeta_3
          + \frac{67\pi^2}{144}
          - \frac{7\pi^4}{240}
   \right) \nonumber \\ &&
+ \left( 
            \frac{200969}{3888}
          + \frac{23\pi^2}{72} \zeta_3
          - \frac{1139}{108} \zeta_3
          + \frac{71}{20} \zeta_5
          + \frac{53\pi^2}{108}
          - \frac{1111\pi^4}{8640} 
\right) \e \nonumber \\ &&
+ \bigg( 
            \frac{4082945}{23328}
          - \frac{11\pi^2}{216} \zeta_3
          - \frac{13109}{162} \zeta_3
          + \frac{901}{36} \zeta_3^2
          - \frac{341}{20} \zeta_5
          + \frac{85\pi^2}{1296}
          - \frac{1943\pi^4}{8640}
          + \frac{257\pi^6}{6720}
\bigg) \e^2 
\Bigg]\nonumber \\
&&+N\,N_F\,\Bigg[ -\frac{7}{12\e^3} - \frac{13}{12\e^2}
+ \frac{1}{\e} \left( \frac{155}{216} +\frac{\pi^2}{72} \right)
\nonumber \\&&
+\left(          - \frac{5905}{1296}
          + \frac{1}{2} \zeta_3
          - \frac{5\pi^2}{72} 
   \right) 
+\left(
          - \frac{162805}{7776}
          - \frac{95}{54}\zeta_3
          - \frac{11\pi^2}{432}
          + \frac{7\pi^4}{1440}
\right) \e\nonumber \\ &&
+\left(
          - \frac{3663205}{46656}
          + \frac{31\pi^2}{108} \zeta_3
          + \frac{274}{81} \zeta_3
          - \frac{9}{10} \zeta_5
          + \frac{883\pi^2}{2592}
          - \frac{73\pi^4}{2592}
\right) \e^2
\Bigg]\nonumber \\
&&+ \frac{N_F}{N} \Bigg[ -\frac{1}{8\e} 
+ \left(\frac{67}{48}-\zeta_3\right)
+\left(     \frac{2027}{288}
          - \frac{23}{6} \zeta_3
          - \frac{7\pi^2}{144}
          - \frac{\pi^4}{54} 
\right) \e\nonumber \\ &&
+\left(            \frac{47491}{1728}
          + \frac{5\pi^2}{18} \zeta_3
          - \frac{281}{18} \zeta_3
          - 4\zeta_5
          - \frac{209\pi^2}{864} 
          - \frac{23\pi^4}{324}
\right) \e^2 
\Bigg]
+ N_F^2\, \frac{1}{9\e^2}
+ {\cal O}(\e^3) 
\;.
\end{eqnarray} 

\section{Conclusions and outlook} 
\label{sec:conc}
In this letter, we computed the two-loop quark and gluon form factors to 
all orders in the dimensional regularisation parameter $\e$. The principal 
ingredient to this calculation is the two-loop crossed triangle graph 
$A_6$, for which we computed an exact expression 
in terms of generalised hypergeometric functions of 
unit argument, which can be expanded to any desired order in $\e$
using the {\tt HypExp}-package.

A potential application of the form factors derived here is the 
extraction of the complete set of
infrared pole terms of the genuine three-loop quark form 
factor from the recently derived three-loop splitting and 
coefficient functions in deep inelastic scattering~\cite{mvv}. In turn, these 
allow to fix the yet unknown hard radiation constants in the 
infrared factorisation formula at three loops. Parts of these constants 
were derived previously from ${\cal N}=4$ supersymmetry 
relations~\cite{kotikov,dixon3l}. 

The two-loop vertex master integrals feature as subtopologies in the 
reduction of the three-loop form factor contributions, appearing if one 
of the three loops is disconnected from the others by pinching  the connecting 
propagators. In this case, their terms to $\e^2$ are required.

The calculation presented here illustrates the applicability of 
the {\tt HypExp}-package in the calculation of multi-loop corrections in 
quantum field theory. Functions similar to those which were expanded here 
appear also in multi-particle phase space integrals in massless~\cite{ggh} 
and massive decay processes~\cite{thuber}. Since all these integrals 
correspond to particular cuts of multi-loop two-point functions, 
one might expect that three-loop and four-loop two-point 
functions could also 
be expanded using {\tt HypExp} to high orders in $\e$, as required for 
multi-loop calculations of fully inclusive observables~\cite{bnew}.

{\bf Note added:} While finalising this letter, an independent 
paper  addressing very similar issues appeared.  
In hep-ph/0507039~\cite{mvvnew}, 
Moch, Vermaseren and Vogt compute the two-loop 
quark form factor to order $\e^2$ and apply it in the extraction 
of the pole parts of the three-loop quark form factor from 
deep inelastic coefficient functions. In this paper, the 
hard radiation constants for infrared factorisation at three-loops and related 
resummation coefficients are extracted for processes involving quarks only. 
Expanding our unrenormalised quark form factor (\ref{eq:fqunren}) to 
order $\e^2$, we confirm the result (B.1) of~\cite{mvvnew}.

\section*{Acknowledgements}
We wish to thank Gudrun Heinrich for independent 
numerical checks of the 
{\tt HypExp}-Laurent expansions using the sector decomposition method 
described in~\cite{secdec}.

This research was supported by the Swiss National Science Foundation
(SNF) under contract 200021-101874.

\end{document}